\documentclass[conference]{IEEEtran}
\IEEEoverridecommandlockouts

\usepackage{balance}

\usepackage{comment}
\usepackage{booktabs}

\usepackage{tabularx}   
\usepackage{booktabs}   
\usepackage{array}      

\usepackage{graphicx}
\usepackage{cite}
\usepackage{amsmath,amssymb,amsfonts}
\usepackage{algorithmicx}
\usepackage{textcomp}
\usepackage{xcolor}

\usepackage{amsthm}

\usepackage[caption=false]{subfig}

\usepackage{authblk}

\newtheorem{theorem}{Theorem}

\usepackage{algorithm,algpseudocode}

\usepackage{multicol}

\usepackage{mathtools}

\usepackage{cuted}

\usepackage{color,soul}
\def\BibTeX{{\rm B\kern-.05em{\sc i\kern-.025em b}\kern-.08em
    T\kern-.1667em\lower.7ex\hbox{E}\kern-.125emX}}

\makeatletter
\newcommand{\linebreakand}{%
\end{@IEEEauthorhalign}
\hfill\mbox{}\par
\mbox{}\hfill\begin{@IEEEauthorhalign}
}

\makeatother

\begin{document}

\title{
Ambiguity Elimination via Successive Estimation in ISAC Systems with Reference Signals

\thanks{This work has been supported by the Smart Networks and Services Joint Undertaking (SNS JU) project 6G-DISAC under the EU’s Horizon Europe research and innovation program under Grant Agreement no. 101139130}
}

\author[1]{Ali Al Khansa}
\author[2]{Sami Mekki}
\author[3]{Giyyarpuram Madhusudan}


\affil[1]{Orange Labs, Rennes, France}
\affil[2]{Nokia Standards, Massy, France}
\affil[3]{Orange Labs, Meylan, France}

\affil[ ]{Emails: \{ali.alkhansa;giyyarpuram.madhusudan\}@orange.com, sami.mekki@nokia.com}

\maketitle

\IEEEpubidadjcol

\begin{abstract}
Integrated sensing and communication (ISAC) systems can efficiently reuse 5G New Radio (NR) reference signals to enable joint communication and passive target detection. Among these, the Positioning Reference Signal (PRS) is widely adopted for high-performance sensing activities. However, the PRS configuration, given by the comb structure, introduces periodic ambiguities, resulting in ghost targets. Existing approaches mitigate this issue by combining PRS with additional reference signals, such as the DeModulation Reference Signal (DMRS), effectively removing ghosts but at the cost of higher complexity and sensing overhead. This paper proposes a low-complexity successive estimation strategy that initially allows ghost targets but progressively eliminates them using historical ambiguity-free estimations. The method comprises a matching process to associate current detections with historical real targets, and a pattern deviation detection mechanism to trigger a fallback to ambiguity-free estimation when the target set changes. We derive a sufficient condition that guarantees the correct matching as a function of target velocity, PRS comb size, subcarrier spacing, and sensing refresh rate. Using this condition, a parameter selection strategy is proposed to jointly improve sensing and communication performance. Numerical results demonstrate that our method achieves unambiguous detection while significantly reducing sensing resources, thus releasing more resources for communication compared to State-of-the-Art (SotA) approaches.

\end{abstract}

\vspace{-0cm}

\begin{IEEEkeywords}
		Integrated Sensing and Communication (ISAC), Orthogonal Frequency Division Multiplexing (OFDM), Ambiguity Elimination, Positioning Reference Signal (PRS).
\end{IEEEkeywords}

\section{Introduction}


6G wireless networks promise to transform the communications landscape by integrating advanced sensing capabilities with conventional data transmission. This Integrated Sensing And Communication (ISAC) paradigm is expected to support various applications, from real-time object tracking, autonomous vehicles \cite{10944644}, environmental monitoring, and smart city implementations \cite{strinati2024distributed}. In particular, the reference signals defined in 5G New Radio (NR) have attracted significant attention for sensing tasks due to their excellent sensing properties, such as good autocorrelation characteristics and robust noise resilience \cite{wei20225g}. Additionally, leveraging these reference signals can lead to cost-effective and simplified system architectures with significant potential for application in radar sensing \cite{lyu2025reference}.



Reference-signal–based sensing in 5G has been widely studied \cite{lyu2025reference}. For example, the work in \cite{wei20225g} compares Positioning Reference Signal (PRS) with other 5G pilots such as DeModulation Reference Signals (DMRS), Channel State Information Reference Signals (CSI-RS), and Synchronization Signals (SS) for range/Doppler estimation, while \cite{huang2022joint} explores joint optimization of channel estimation, detection, and pilot selection. Among these, PRS (introduced in 3GPP Rel-16) stands out for its rich time–frequency resources and flexible configuration. 
Its well-designed pseudo-random sequence structure provides a strong correlation between the transmitted and received signals, and \cite{wei20225g} reports superior sensing performance relative to other pilots. 
However, since the PRS was originally designed for positioning, its sparse Resource Element (RE) pattern causes periodic ambiguities in passive sensing applications, resulting in ghost targets that compromise reliable detection.

Recent works have addressed this issue using different methods. For example, \cite{wei2024multiple} proposes a solution that uses multiple reference signals together to resolve the ambiguity. Under the same umbrella of leveraging multiple reference signals, the works of \cite{khosroshahi2024leveraging} and \cite{khosroshahi2024doppler} address ambiguity issues using a 3GPP-compliant solution that jointly employs DMRS and PRS to remove range ambiguity \cite{khosroshahi2024leveraging} and Doppler ambiguity \cite{khosroshahi2024doppler}. Other methods propose irregular resource patterns to eliminate the ambiguity, as in \cite{golzadeh2024joint}, where irregular time-frequency domain solutions are proposed. 
However, current literature has some limitations. As mentioned in \cite{wei2024multiple}, the flexibility of employing multiple reference signals introduces system design complexity, as the parameters must be optimized to accommodate the continuous distribution of these signals. This challenge is evident in the solutions of \cite{wei2024multiple,khosroshahi2024leveraging,khosroshahi2024doppler}. Another issue is the compatibility of the solutions with the reference signal standards, such as when irregular resource patterns are used as proposed in \cite{golzadeh2024joint}. In addition, irregular pattern methods should be considered wisely, as reducing the ambiguity through irregular resource patterns could come at the cost of increasing side lobes attenuation \cite{hakobyan2019ofdm}. Moreover, resource allocation between sensing and communication and tradeoff analysis are also concerned \cite{al2024dynamic,al2025adaptive,al2025learning}, where the sensing overhead in the State-of-the-Art (SotA) solutions aimed at ambiguity reduction is often increased, e.g., due to allocating more sensing resources, thereby limiting the communication resources.

In this paper, we address the problem of the ambiguity induced by the empty subcarriers in comb-structured reference signals (e.g., PRS). However, rather than relying continuously on multiple reference signals, we propose a low-complexity approach that minimizes this dependence through successive estimation. Our method initiates the algorithm with an ambiguity-free estimation using the existing solutions in SotA (e.g., via multiple reference signals). Then, the resulting target estimates are stored as a historical reference. In subsequent estimations, a matching process compares the current estimates with the historical data to distinguish genuine targets from ghost artifacts. Additionally, we derive sufficient conditions relating target velocity, PRS allocation pattern defined by the comb size $K_{\text{comb}}$, subcarrier spacing, and sensing refresh rate, which guarantee reliable target matching. These conditions are used to optimize resource allocation by enabling larger $K_{\text{comb}}$ and longer refresh intervals when the target scene is stable, thereby reducing the sensing overhead and improving communication efficiency without sensing ambiguity problems. 
Numerical results demonstrate that our proposed approach not only effectively mitigates ambiguity but also dynamically adapts resource allocation in response to changes in the target scene, which reduces sensing overhead and enables an efficient and scalable ISAC system.

The rest of the paper is structured as follows: Section II introduces the system model, the successive estimator (matching process and pattern deviation strategy) is detailed in Section III, Section IV provides the numerical results, and Section V concludes the paper.
\section{ISAC system Model}
\subsection{OFDM Radar Fundamentals}
\begin{figure}[t!]
\centering
\includegraphics[scale=0.2]{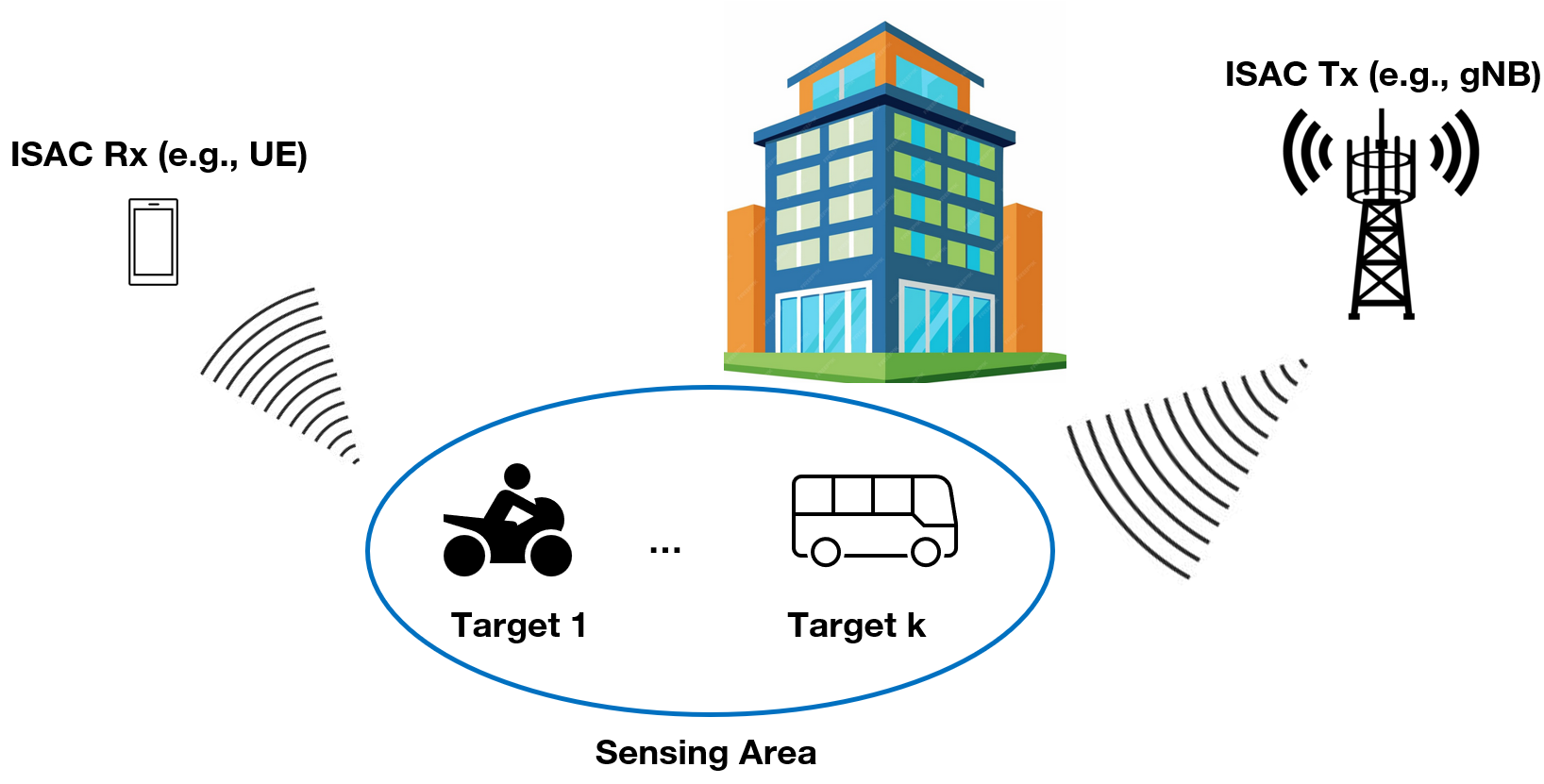}
\caption{ISAC system model.}

\label{fig:SystemModel}
\end{figure}

In this work, we consider a downlink scenario in which a gNB transmits signals over a sensing region containing $K$ point-like targets. The receiver captures the backscattered echoes from these targets to estimate their range and velocity. We assume that all targets lie in the line-of-sight of both the transmitter and the receiver. Figure \ref{fig:SystemModel} illustrates an example of the considered scenario. 

Let $N$ be the number of OFDM symbols in the time domain and $M$ be the number of subcarriers in the frequency domain. The continuous-time transmitted signal $s(t)$ is written as \cite{khosroshahi2024leveraging}:
\begin{equation} 
s(t) = \sum_{n=0}^{N-1} \mathrm{rect}\Bigl(\tfrac{t - nT_{0}}{T_{0}}\Bigr)\sum_{m=0}^{M-1} v(m,n) e^{j 2\pi m \Delta f(t - nT_{0})},
\end{equation}
where $\mathrm{rect}(\cdot)$ denotes the rectangular pulse, $\Delta f$ is the subcarrier spacing, and $T_0 = T_{\text{CP}} + T_s$ is the total OFDM symbol duration (with $T_s= 1/\Delta f$ and $T_{\text{CP}}$ the cyclic prefix length). The term $v(m,n)$ is the complex data symbol assigned to subcarrier $m$ of OFDM symbol $n$.

Suppose $K$ point-like targets are present in the sensing region. The received signal $r(t)$ is then modeled as \cite{braun2014ofdm}:
\begin{equation} 
r(t) = \sum_{k=1}^{K} \beta_{k}s\bigl(t - \tau_{k}\bigr)e^{j2\pi f_{d,k}t} + u(t),
\end{equation}
where $\beta_{k}$ is the attenuation factor of the $k$-th target, $\tau_{k}$ is the propagation delay, $f_{d,k}$ is the Doppler frequency shift, and $u(t)$ is the complex Additive White Gaussian Noise (AWGN) term with zero mean and variance $2\sigma^2$.

In conventional OFDM-based communication systems, the receiver samples each OFDM symbol and applies a Fast Fourier Transform (FFT) to retrieve the transmitted symbols. Under typical conditions, the cyclic prefix $T_{\text{CP}}$ absorbs channel delays, and Doppler effects remain small, thereby minimizing Inter-Symbol Interference (ISI). However, in high-mobility or long-range sensing scenarios, the path delay and Doppler shift can exceed the guard interval, causing ISI in both the time and frequency domains. Consequently, the received sample on subcarrier $m$ and OFDM symbol $n$ may be approximated by: 
\begin{equation}
\begin{aligned} 
y(m,n) & = \sum_{k=1}^{K} \beta_{k}e^{j2\pi nT_{0}f_{d,k}}e^{-j2\pi m\Delta f \tau_{k}}v(m,n) + \mathrm{ISI} \\
       & + q(m,n),
\end{aligned} 
\end{equation}
where $q(m,n)$ is the complex AWGN term with zero mean and variance $2\sigma^2$. The ISI component arises from uncompensated delay and Doppler shifts, which may distort the received symbols in challenging sensing conditions.

\subsection{Periodogram-Based Estimation }
It is well known that the received signal samples experience phase shifts due to Doppler effects on the time axis and due to time shifts on the frequency axis. Estimating the range and Doppler requires a two-dimensional search, for which multiple strategies are available. Several techniques are discussed in \cite{braun2014ofdm}, but we focus on the periodogram method motivated by its low computational complexity. Due to size limitations, we refer readers to \cite{khosroshahi2024leveraging} for the detailed equations, and we briefly present the significant steps of this process for completeness.

The periodogram operates in two stages: first, the range is estimated, followed by Doppler estimation after delay compensation. The received signal $y$ is divided element-wise by the data $v$ to obtain $g$. The magnitude of the $M$-point Inverse FFT (IFFT) is then computed for each column of $g$,  yielding $r_n(l)$, where $l\in\{0,\dots,M-1\}$. Averaging across $N$ symbols improves the SNR, resulting in \mbox{$\overline{r}(l)\triangleq \tfrac{1}{N}\sum_{n=0}^{N-1} r_n(l)$}. The peak index $\widehat{l}_k=\arg\max_l \overline{r}(l)$ maps to the bistatic range via $\widehat{R}^{\mathrm{tot}}_{k}=\frac{c\,\widehat{l}_k}{\Delta f\,M}.$ Next, delay compensation is applied for each detected target and a Doppler profile is formed by taking the magnitude of the $N$-point FFT across the rows of $g$, yielding $v_m(d), d \in \{0,\dots, N-1\}$. Averaging across $M$ subcarriers gives $\overline{v}(d)\triangleq \tfrac{1}{M}\sum_{m=0}^{M-1} v_m(d)$. Finally, the peak index $\widehat{d}_k=\arg\max_d \overline{v}(d)$ is mapped to the velocity as $\widehat{v}=\frac{c\,\widehat{d}_k}{2\,T_s\,f_c\,N}$.

\section{Ghost Targets Elimination via successive estimation}

\subsection{Ghost Targets and How to Use DMRS to Remove Them}
\begin{figure}[t!]
\centering
\includegraphics[scale=0.40]{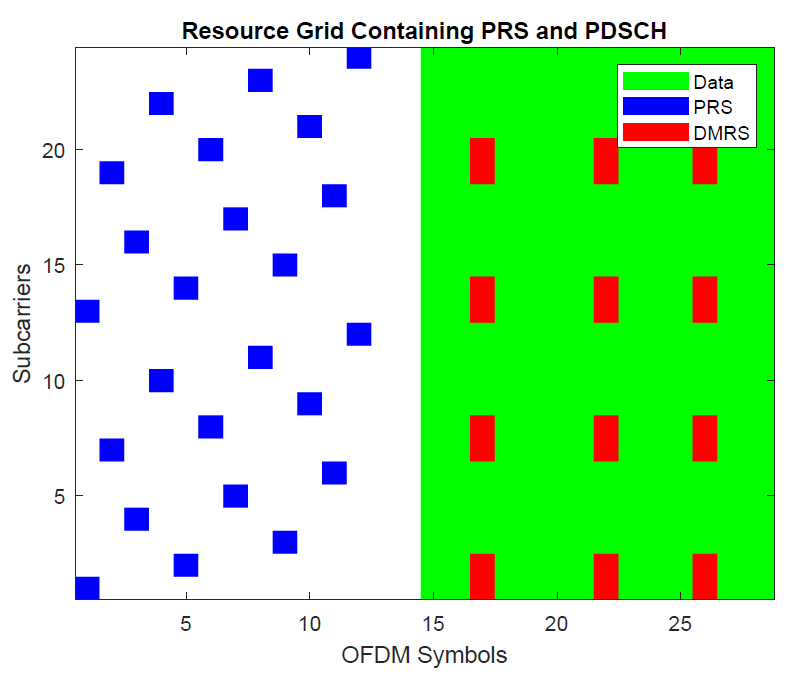}
\caption{PRS–PDSCH resource allocation: 2 slots, 2 PRBs, comb size 12 \cite{khosroshahi2024leveraging}.}
\label{PRS&DMRS}
\end{figure}

When used for passive target sensing, the comb-structured allocation of the PRS in 5G NR introduces empty subcarriers that cause periodic ambiguities in the range and Doppler domains. These manifest as ghost targets, which are replicas of the real target responses shifted by fixed intervals. A 3GPP-compliant method to remove ghost targets has been proposed in \cite{khosroshahi2024leveraging}, where the DMRS embedded in the Physical Downlink Shared Channel (PDSCH) is repurposed for sensing alongside PRS. Figure \ref{PRS&DMRS} illustrates an example of such resource allocation in the time domain between PRS and PDSCH, and shows an example of the comb-structure of PRS where  $K_{\text{comb}}=12$. 

The main idea is to exploit the fact that PRS and DMRS have different ghost patterns. By combining their outputs appropriately, ghost peaks that appear in one but not the other are suppressed, leaving only the real targets. In this approach, PRS is first processed using standard range estimation to produce a range profile, and the same is done for DMRS extracted from PDSCH. Different fusion strategies can be applied \cite{khosroshahi2024leveraging,khosroshahi2024doppler}, and below are some examples of how to remove range ambiguities \cite{khosroshahi2024leveraging} for $K_{\text{comb}} \in\{2,4,6,12\}$.
\begin{itemize}
    \item For smaller comb sizes (2 or 4), the ghost patterns of PRS and DMRS are sufficiently distinct. Here, the range profiles from the two signals are combined using a Hadamard product. Since the ghost in one profile typically corresponds to a low value in the other, this product cancels ghosts while preserving real peaks.
    \item For larger comb sizes (6 or 12), some ghost peaks overlap between PRS and DMRS, so direct multiplication is less effective. In this case, the method sums the time-domain representations of PRS and DMRS, thereby disturbing the periodicity of the ghost pattern and reducing the magnitude of ghost peaks below detection thresholds.
\end{itemize}
This solution maintains range resolution and maximum detection range and is fully compatible with the 5G NR specifications, as it does not require changes to gNB configurations. The two cases (PRS-only and PRS+DMRS) are illustrated in Figure \ref{fig:OldResults} and will be discussed in the numerical results section. For Doppler ambiguities, a similar method can be applied \cite{khosroshahi2024doppler}. These methods will serve as the initialization step of our proposed algorithm, as detailed in the following subsections.

While effective, the SotA approaches inherit the drawbacks of multi-reference-signal schemes noted in the introduction: increased complexity, dependence on smaller $K_{\text{comb}}$ values to avoid performance degradation, and higher sensing overhead from allocating multiple reference signals. These limitations motivate the development of lower-complexity alternatives. 

\subsection{Successive Estimation}
\label{subsectionSuccEst}
To address the mentioned limitations, we propose a simple and practical scheme that resolves ambiguities and false positives (ghost targets) while minimizing reliance on multiple reference signals. The core idea is to offer a low-complexity solution that initially tolerates the presence of ghost targets but progressively eliminates them through successive estimation. This approach leverages unambiguous prior estimations to disambiguate subsequent detections. Our proposal is built upon two key observations:
\begin{enumerate}
\item Ghost targets appear in periodic patterns, which can be exploited to assist ambiguity elimination.
\item Successive estimations provide side information about genuine targets, which can be matched to current detections to filter out ghost targets.
\end{enumerate}

Successive estimation involves storing the range and Doppler estimates obtained from previous ambiguity-free measurements and using them in future estimations to distinguish real from false targets. While the strategy can be extended to both range and Doppler, we focus here on range estimation without loss of generality. The proposed method consists of two main components:
\begin{itemize}
\item \textit{The Matching process}: For each target in the historical data, we select the closest detected peak in the current measurement, treating remaining peaks as likely ghosts.
\item \textit{The Pattern deviation detection}: A mechanism to detect inconsistencies in the periodic structure of ghost peaks, which serves as a trigger to re-invoke a full ambiguity-free estimation using SotA methods.
\end{itemize}

\subsection*{The Matching Process}
The goal of the matching process is to associate currently detected targets with previously known (historical) ones. This is done by selecting, for each historical range estimate, the closest peak from the current detection. The rest of the peaks are considered likely ghost targets. Assume there are $x$ real targets with historical range values $r_{\text{history},i}$ for $i\in\{1,\dots,x\}$  and $y$ total targets (real + fake) with detected ranges $r_j, j\in\{1,\dots,y\}$. The current range estimate for each target $r_{\text{real},i}$ is selected by solving:
\begin{equation}
    r_{\text{real},i} = r_{\widehat{j}},  \forall i \in \{1, \dots, x\},
    \label{MatchingEquation}
\end{equation}
with
\begin{equation}
    \quad \widehat{j} = \arg \min_{j \in \{1, \dots, y\}} | r_{\text{history},i} - r_j |, \forall j \in \{1, \dots, y\}.
\end{equation}
This formulation ensures that each historical target is matched to the closest candidate in the current estimation.

\subsection*{The Pattern Deviation Detection}
The second component of our strategy determines whether to rely on successive estimation or to re-trigger a full ambiguity-free estimation using SotA methods. Since newly appearing targets are not represented in the historical record, they require a fresh, unambiguous estimation. To detect such events, we exploit the inherent periodicity of ghost targets. The distance between ghost peaks calculated from the two-way propagation delay is given by \cite{golzadeh2024joint}:
\begin{equation}
    d_{\text{ghost}} = \frac{c}{2 K_{\text{comb}} \Delta f}.
\label{dghost}
\end{equation}
Table \ref{table1} shows how $d_{\text{ghost}}$ varies depending on the Sub Carrier Spacing (SCS) (i.e., $\Delta f$) and $K_{\text{comb}}$. For one-way propagation delay, the ambiguity intervals are twice those in Table \ref{table1}.
\begin{table}[H]
    \centering
    \caption{Range interval between ambiguities (two-way) [in meters] }
    \begin{tabular}{lcccc}
        \toprule
        & \textbf{comb 2} & \textbf{comb 4} & \textbf{comb 6} & \textbf{comb 12} \\
        \midrule
        \textbf{SCS = 15 kHz}  & 4996 & 2498 & 1665 & 832 \\
        \textbf{SCS = 30 KHz}  & 2498 & 1249 & 832 & 416 \\
        \textbf{SCS = 60 kHz}  & 1250 & 624 & 416 & 208 \\
        \textbf{SCS = 120 KHz} & 624 & 312 & 208 & 104 \\
        \textbf{SCS = 240 kHz} & 312 & 156 & 104 & 52 \\
        \bottomrule
    \end{tabular}
    
    \label{table1}
\end{table} 




Using $d_{\text{ghost}}$, we can predict the expected positions of ghost targets for each real target. After performing the matching process, the matched and the unmatched peaks are analyzed. For the matched peaks, we check how far they are from the history, and given a tolerance value (e.g., tolerance based on the velocity of the targets), we check if the matching process is okay or not. For the unmatched peaks, say $r_{\text{ghost}}$, we compute its normalized offset from the real targets:
\begin{equation}
\text{OffsetRatio}_i =  \left( \frac{ |r_{\text{ghost}} - r_{\text{real},i}| }{ 2\cdot d_{\text{ghost}} } \right), i\in\{1,\dots, x\}.
\end{equation}
If for a given real target $i\in \{1,\dots,x\}$, the offset ratio $\text{OffsetRatio}_i$ is sufficiently close to an integer (e.g., within a predefined threshold $\epsilon$), the peak is classified as a ghost. Otherwise, it indicates the presence of a new real target, and a pattern deviation is declared. This triggers a new round of ambiguity-free estimation. An illustration of the different results of the matching process is seen in the numerical section (check Figure \ref{fig:NR_MatchingMerged}). One edge case may arise when a real target happens to appear at a location indistinguishable from a ghost, \textit{i.e.}, offset ratio close to an integer. In such cases, periodicity-based detection may fail. However, a significant and sustained change in peak magnitude across successive estimations can provide an auxiliary indication to detect this scenario.


\subsection{Sufficient Conditions of the Matching Process}
Following the method detailed in \ref{subsectionSuccEst}, we now examine the practical conditions under which the proposed matching strategy reliably identifies real targets. Assuming no pattern deviation is detected (i.e., ghost peaks appear as expected), the matching process may still result in an error if a target moves significantly between two sensing intervals. In particular, a mismatch may occur when a real target shifts far enough toward the location of one of its ghost replicas. In such cases, the matching algorithm may incorrectly associate the target with a previous position rather than its new shifted location.

The likelihood of such errors depends on several factors: the distance between ghost peaks ($d_{\text{ghost}}$), the velocity of the target, the direction of movement, and the sensing refresh rate, denoted $\Delta t$. Intuitively, the longer the system waits before refreshing the sensing data, or the faster the target moves, the higher the risk of mismatch. To address this, we define a worst-case scenario and derive a sufficient condition on the maximum allowable target velocity that ensures reliable matching. This condition enables us to select optimal values for $K_{\text{comb}}$ and the sensing refresh rate $\Delta t$, two critical degrees of freedom in balancing the efficiency of the ISAC system.


\begin{theorem}\label{th1}
Let $v$ denote the target’s velocity. The matching process correctly identifies the real targets if:
\begin{equation}
\Delta t K_{\text{comb}} < \frac{c}{4 v \Delta f}.
\label{SufficientCondition}
\end{equation}
\end{theorem}

\begin{proof}
Over one sensing interval $\Delta t$, the worst-case two-way range displacement is
\begin{equation}
\Delta d \;\le\; 2\,v\,\Delta t.
\end{equation}
Reliable matching requires the displacement to be smaller than half of the distance between adjacent ghost peaks (two-way):
\begin{equation}
2\,v\,\Delta t \;<\; d_{\text{ghost}}.
\end{equation}
Substituting $d_{\text{ghost}}$ and rearranging yields \eqref{SufficientCondition}.
\end{proof}
Theorem \ref{th1} expression offers a practical design guideline. For a given maximum target velocity $v$ and subcarrier spacing $\Delta f$, the system designer can jointly choose the largest possible values of $K_{\text{comb}}$ and $\Delta t$ that satisfy the inequality. Larger values are preferred as they reduce the frequency of sensing operations and the reference signal overhead, thus achieving better resource utilization. An illustration on the optimal allocation is seen in the numerical section (check Figure \ref{fig:OptimalPairs}).

\subsection{Advantages of the Proposal}
A key advantage of the proposed successive estimation approach is its potential to significantly reduce sensing overhead. By enabling the use of higher values for both $K_{\text{comb}}$ and $\Delta t$ when pattern deviation is not detected, the system can allocate more resources to communication and reduce the frequency of sensing operations. However, the actual resource savings depend on how frequently the system must fall back to the SotA solution. This, in turn, is influenced by the dynamic nature of the sensing environment, particularly the rate at which targets appear or disappear. Let $p$ denote the probability of encountering a pattern deviation event, \textit{i.e.} the probability that a reversion to the SotA ambiguity-free estimation is triggered. Assuming the SotA baseline uses fixed values of $K_{\text{comb}}$ and $\Delta t$, and that our method alternates between this baseline (with probability $p$) and a more efficient configuration using $\max(K_{\text{comb}})$ and $\max(\Delta t)$ (with probability $1-p$), the relative gain in resource efficiency can be expressed as:
\begin{equation}
\text{Gain} = \left[1 - \frac{\left( \frac{p}{K_{\text{comb}}} + \frac{1-p}{\max(K_{\text{comb}})} \right) \cdot \frac{1}{\max(\Delta t)}}
{\frac{1}{K_{\text{comb}}} \cdot \frac{1}{\Delta t}} \right] \times 100.
\label{GainEq}
\end{equation}
This expression compares the average number of sensing resource slots used per unit time in our method versus the SotA baseline. A higher gain indicates better resource utilization and more capacity left for communication functions. An illustration of the effect of the different parameters on this gain is presented in the numerical results (check Figure \ref{fig:Gain}). That said, the complete process is presented in Algorithm (\ref{alg:range_estimation}). Note that $\texttt{Pattern\_Deviation()}$ is a simple Boolean function that returns 1 (true) if there is a pattern deviation detection.

\begin{algorithm}[h]
\caption{Proposed Successive Estimation Algorithm}
\label{alg:range_estimation}
\begin{algorithmic}[1]

\State \textbf{Initialize:} $\textit{Modified} \gets 0$
\State \textbf{Perform initial range estimation} using ambiguity-free methods from SotA. 
\State Compute range $r_i$ for each detected target $i \in \{1, \dots, x\}$.
\State Set $r_{\text{history}_i} \gets r_i$ for all $i \in \{1, \dots, x\}$ (update history).
\State \textbf{Optimize $K_{\text{comb}}$ and $\Delta t$} according to the mobility scenario (while satisfying Eq. (\ref{SufficientCondition})).
\While{(\textit{Modified} $= 0$)}
    \State \textbf{Re-estimate $r_i$} using PRS only.
    \State Match the $x$ real ranges using the history values via the closest matching method for all $r_i$ (as in Eq. (\ref{MatchingEquation})).
    \State Update history: $r_{\text{history}_i} \gets r_i$ for all $i \in \{1, \dots, x\}$.
    \If{(\texttt{Pattern\_Deviation()=1})}
        \State \textit{Modified} $\gets 1$        
    \EndIf
\EndWhile
\State \textbf{Repeat:} Go to step 2 (new unambiguous estimation call).
\end{algorithmic}
\end{algorithm}

\section{Numerical Results}

We validate our proposal  via simulations using MATLAB 5G toolboxes, ensuring the approach's compatibility with 3GPP standards. The setup involves a gNB transmitting a 5G NR signal with a carrier frequency of 28 GHz and an SCS of 120 kHz towards a sensing area containing two passive targets. The targets are located at bistatic distances of 1057 m and 1242 m, with SNR = 0 dB. The PRS is configured to occupy 12 symbols. The DMRS uses configuration type 2, with three DMRS symbols located at symbol positions {3, 8, 12} within each slot. Additionally, the PDSCH employs mapping type "A". The code rate used to determine the transport block sizes is set to 490/1024, as specified in \cite{3gpp2020nr}. The Virtual resource block (VRB) bundle size is 4, and VRB to PRB interleaving is disabled, and we consider the 16 Quadrature Amplitude Modulation (16-QAM) for the PDSCH.

\begin{figure}[t!]
\vspace{-0 pt}
\centering
\includegraphics[scale=0.59]{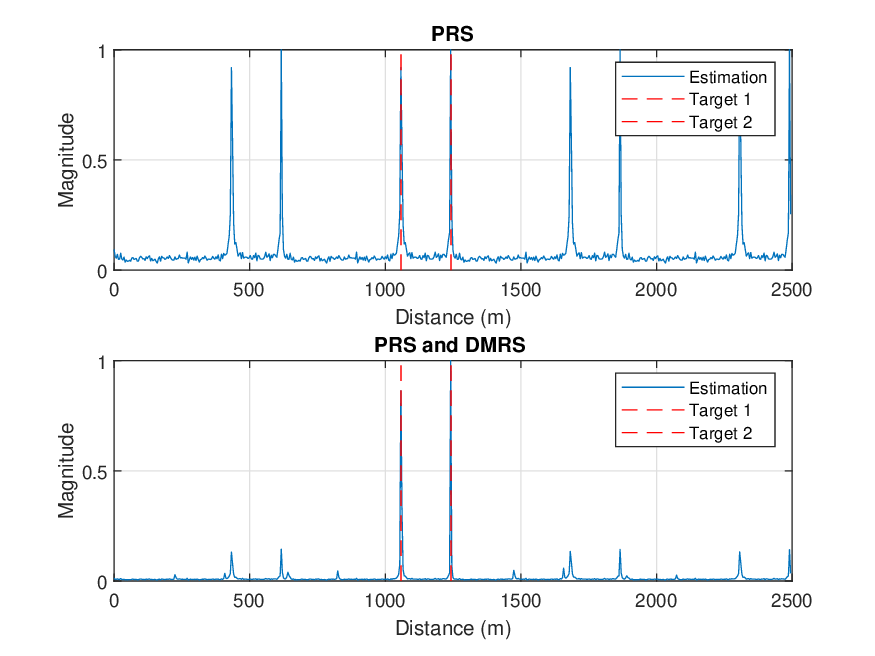}

\caption{Range estimation using PRS only (top) and PRS combined with DMRS (bottom) for a two-target scenario ($ K_{\text{comb}} = 4, fc = 28$GHz and $\text{SCS}=120$kHz). Ghosts are eliminated when combining PRS with DMRS.}
\label{fig:OldResults}
\end{figure}

\begin{figure}[t!]
\vspace{-0 pt}
\centering
\includegraphics[scale=0.59]{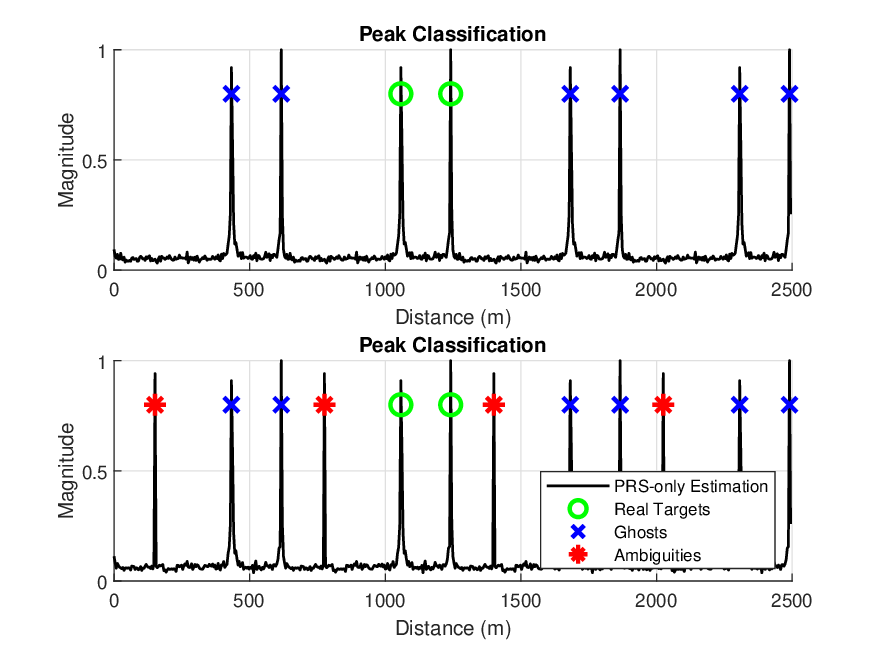}

\caption{The proposed matching process in two scenarios: correct matching with no pattern deviation (top) and ambiguity detection due to pattern deviation (bottom). A third target is detected, which modified the pattern.}
\label{fig:NR_MatchingMerged}
\end{figure}

We present four figures to evaluate and illustrate the behavior of the proposed successive estimation method, compare it to the SotA solution, and demonstrate its practical efficiency in terms of sensing resource savings. Figure \ref{fig:OldResults} compares the performance of range estimation using PRS only (top) versus a combination of PRS and DMRS (bottom). As shown in the top subplot, the PRS-only configuration results in significant ambiguities, with multiple ghost peaks periodically appearing around the actual targets. This behavior stems from the comb-structured PRS allocation, which leaves unoccupied subcarriers and introduces aliasing in the range domain. In contrast, the bottom subplot shows the result of using PRS and DMRS together, which effectively eliminates the ambiguities. The target detection is clean, and only the real targets remain visible. This demonstrates the strength of existing SotA ambiguity elimination techniques and justifies our use of such methods for the initialization of our proposed framework.

Figure \ref{fig:NR_MatchingMerged} demonstrates how the proposed successive estimation algorithm performs in different scenarios. In the top subplot, the matching process successfully identifies the real targets based on historical values, as the ghost peaks appear exactly at distances consistent with the theoretical $d_{\text{ghost}}$ periodicity. This confirms that no pattern deviation is detected, and the system can rely on successive estimation. In contrast, the bottom subplot shows a scenario where a new target has entered the sensing field, generating ghost peaks that no longer align with the expected periodic structure. The deviation is detected by our pattern check, triggering a switch to an ambiguity-free estimation in the next cycle. These results show the critical role of the deviation detection strategy in maintaining accuracy over time without unnecessary overhead.

\begin{figure}[t!]

\centering 
\includegraphics[scale=0.59]{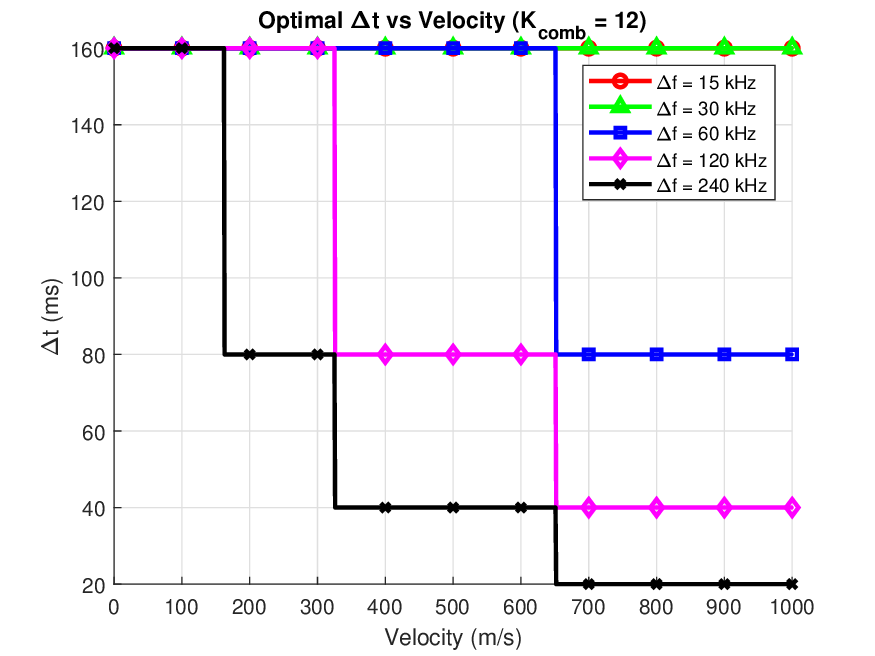}
\caption{Optimal sensing refresh rate $\Delta t$ versus target velocity for $K_{\text{comb}}=12$ and different $\Delta f$, satisfying the sufficient condition Eq. (\ref{SufficientCondition}) $(\Delta t \in \{20,40,80,160\}\text{ ms})$.}
\label{fig:OptimalPairs}
\end{figure}

Figure \ref{fig:OptimalPairs} explores the practical mobility constraints for ensuring correct operation of the matching process. Using the sufficient condition in Theorem \ref{th1}, we compute the optimal sensing refresh rate $\Delta t$ for a given maximum velocity, assuming $K_{\text{comb}} = 12$, which is found to be the most resource-efficient setting in the practical range of target velocities. Figure \ref{fig:OptimalPairs} presents the optimal $\Delta t$ versus velocity for different subcarrier spacings $\Delta f$. As expected, higher target velocities require more frequent sensing updates (i.e., smaller $\Delta t$) to maintain matching accuracy. In practice, systems usually operate with a fixed set of $\Delta t$ values; here we assume $\Delta t \in \{20, 40, 80, 160\}$ ms. The curves in Figure \ref{fig:OptimalPairs} provide system designers with a reference to select the most relaxed (i.e., largest) $\Delta t$ that satisfies the sufficient condition for a given velocity and $\Delta f$.

\begin{figure}[t!]

\centering
\includegraphics[scale=0.59]{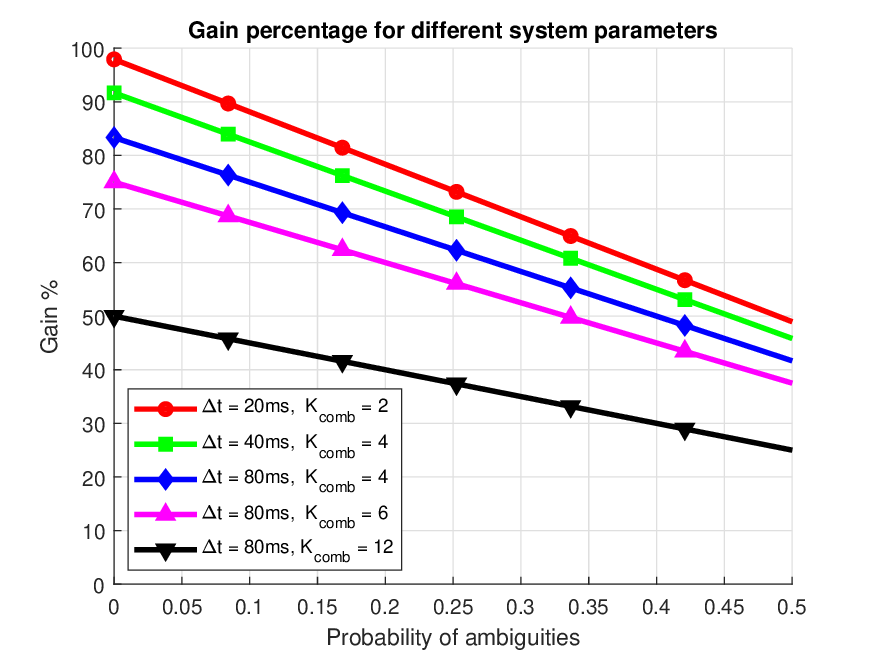}
\caption{Percentage gain (Eq. (\ref{GainEq})) in sensing resource usage of the proposed method compared to the SotA baseline, as a function of the probability $p$ of pattern deviation events and different SotA $\Delta t$ and $K_{\text{comb}}$ parameters.}
\label{fig:Gain}
\end{figure}

Finally, Figure \ref{fig:Gain} quantifies the sensing resource savings achieved by the proposed method. We plot the percentage gain in resource usage compared to a SotA baseline that uses a fixed $(K_{\text{comb}}, \Delta t)$ pair. The x-axis represents the probability $p$ of pattern deviation events, which dictate how often the system must revert to the more costly SotA strategy. The results show that for small $p$ (i.e., when scene changes are infrequent), the gain is substantial, reaching over $80\%$ in some configurations. As $p$ increases, the gain naturally diminishes, reflecting the increased reliance on full-reference methods. Additionally, we observe that the gain is highest when the SotA baseline uses smaller $K_{\text{comb}}$ and shorter $\Delta t$ values, since the resource cost of such settings is higher, leaving more room for improvement through our adaptive approach.

\section{Conclusions and Future Works}
This paper proposed a practical successive estimation strategy for ghost target elimination within ISAC systems based on reference signals. Unlike the SotA approaches, the proposed method uses historical ambiguity-free estimations to progressively remove ghost peaks in subsequent sensing periods. Two key components were introduced: a matching process to associate current detections with historical real targets, and a pattern deviation detection mechanism to identify when a fallback to an ambiguity-free estimation is required. A sufficient condition was derived to guarantee correct matching, enabling the selection of optimal sensing parameters for a given mobility scenario. Numerical results demonstrated that the proposed method can significantly reduce sensing overhead and free resources for communication, while maintaining unambiguous detection performance close to that of the SotA.

While this work focused on line-of-sight scenarios and numerical evaluations, several extensions naturally follow. In particular, the proposed successive estimation framework is not inherently limited to line-of-sight conditions. In non-line-of-sight environments, the ambiguity periodicity induced by comb-structured reference signals remains unchanged, while additional peaks arise due to multipath propagation. An extension of the proposed matching and pattern deviation mechanisms to operate on multipath additional peaks constitutes an interesting direction for future work. Moreover, future work will target the issue of initialization step, i.e., generalize the work to use only PRS even at the initialization step.  

From a theoretical and complexity perspective, this work establishes a sufficient condition that guarantees correct target matching as a function of target velocity, PRS comb size, subcarrier spacing, and sensing refresh rate. This result provides a practical design guideline for adaptive ISAC parameter selection. While the numerical results in this paper focus on demonstrating the feasibility and resource-efficiency gains of the proposed approach, further investigations could include probability-of-detection and false-alarm analyses under varying SNR and mobility conditions. Moreover, the proposed method significantly reduces sensing overhead as well as the overhead of the channel estimate since DMRS is not often used. By relying predominantly on PRS-only estimation and invoking ambiguity-free processing only upon detected pattern deviations, the approach lowers signaling and processing demands while remaining fully compliant with 3GPP specifications.

\bibliography{conf1}

\end{document}